\documentclass[fleqn,11pt,twoside]{article}
\usepackage{espcrc1}
\usepackage{epsfig,amssymb}
\long\def\@makefigurecaption#1#2{#1. #2\par}

\newcommand{\AmS}{{\protect\the\textfont2
  A\kern-.1667em\lower.5ex\hbox{M}\kern-.125emS}}
\title{Preliminary Results on $\gamma\gamma \to K_sK^\pm\pi^\mp$ from
 $e^+e^-$ Scattering at CLEO}
 \author{V.~Savinov\thanks{Representing CLEO Collaboration, corresponding author, e-mail: vps3@pitt.edu.}
and R.~Ahohe, \addressmark{Department of Physics and Astronomy, \\
University of Pittsburgh, 3941 O'Hara St.,
Pittsburgh, PA, 15241, USA}
 }

\begin{document}
\maketitle

\begin{abstract}
We analyzed 13.8 ${\rm fb^{-1}}$ of the integrated $e^+e^-$ 
luminosity collected at 10.6 GeV center-of-mass energy with the 
CLEO II and II.V detectors to study exclusive two-photon 
production of single hadronic resonances. We searched for hadrons 
decaying into $K_s K^\pm \pi^\mp$ when both leptons 
remain undetected. In this analysis we studied the detection efficiency 
and evaluated systematic errors using independent 
data samples. We estimated 90\% CL upper limits on the products of the two-photon 
partial widths of (pseudo)scalar hadrons with masses below 1.7 ${\rm GeV/c^2}$ 
and their branching fractions into $K_s(\pi^+\pi^-)K^\pm\pi^\mp$. 
Our preliminary results are marginally consistent with the first 
observation of $\eta(1440)$ in two-photon collisions by the L3 experiment. 
\end{abstract}

\section{Introduction}
\label{introduction}
A key to understanding the phenomenon of quark and gluon confinement in
Quantum Chromodynamics (QCD) 
would be an experimental observation and the analysis of the properties 
of unusual hadronic bound states predicted by Lattice QCD (LQCD),  
Flux Tube and other models. 
For example, the Flux Tube model and LQCD (even when the quenching approximation is lifted) 
predict a large number of light glueballs -- bound states of the carriers 
of strong interaction and hybrids -- hadrons composed of three constituents, 
two quarks and a gluon. 
A large number of proposed candidates for these new states of matter
have been observed over the past 35 years in many experiments\cite{PDG}.
In our opinion, most of these candidates need to be confirmed and remain
to be understood. One such candidate is the infamous $\eta(1440)$
first observed in 1967 in $p\bar{p}$ annihilation at rest into $K\bar{K}\pi\pi^+\pi^-$.
This resonance has also been sighted in radiative decays of $J/\psi$
into $K\bar{K}\pi$ and in charge-exchange hadronic reactions $\pi^-p \to \eta \pi\pi n$.
Until recently $\eta(1440)$ has been observed only in gluon-rich environments 
and this established it as a prominent glueball candidate. 
Another hypothesis for $\eta(1440)$'s internal structure is that
it is simply a radially excited $\eta$ meson.

One way to discriminate 
the ground state meson, radial excitation, and glueball hypotheses
is to measure the two-photon partial width of $\eta(1440)$. 
Assuming that quantum numbers allow a two-photon 
decay, its partial width would be of a keV order for a ground state meson, 
approximately an order of magnitude smaller for a radial excitation\cite{page}, and 
of a vanishingly small value for a true glueball because photons couple to gluons 
only through an intermediate quark loop. Finally, it is also possible 
that light glueballs and mesons are mixed and their parameters 
should be obtained from global fits to light hadrons spectrum, 
which recently became an interesting research topic in its own right. 
In 2001 a new piece was added to the $\eta(1440)$ puzzle when
the L3 experiment reported\cite{L3} the first observation of the $\eta(1440)$
in two-photon collisions and measured its two-photon partial width
to be $212 \pm 50~{\rm (stat.)}\pm 23~{\rm (sys.)}$ eV, assuming 100\% branching fraction to
$K\bar{K}\pi$. In our analysis we tried to verify the claim made by L3 
and to measure the two-photon partial width of $\eta(1440)$ with better precision. 

\section{Experimental Apparatus, Data Sample and the Analysis Procedure}
\label{apparatus}
The results presented here
were obtained from
the data accumulated at the
Cornell Electron Storage Ring (CESR) 
with the CLEO series of detectors.
These results are based on statistics that correspond
to the integrated $e^+e^-$ luminosity of
$13.8 {\rm ~fb^{-1}}$
collected at and 60 MeV below the $\Upsilon(4{\rm S})$ energy.
The first third of the data was recorded with
the CLEO II detector\cite{CLEOII_description}
which consisted of three cylindrical drift chambers
placed in an axial solenoidal magnetic field of 1.5T,
a CsI(Tl)-crystal electromagnetic calorimeter,
a time-of-flight (TOF) plastic scintillator system
and a muon system (proportional counters embedded
at various depths in the steel absorber).
Two thirds of the data were taken with
the CLEO II.V configuration of the detector
where the innermost drift chamber
was replaced by a silicon vertex detector\cite{CLEOII.V_description} (SVX)
and the argon-ethane gas of the main drift chamber
was changed to a helium-propane mixture.
This upgrade led to improved resolutions in momentum
and specific ionization energy loss
($dE/dx$) measurements.

The information from the two outer drift chambers,
the TOF system and electromagnetic calorimeter 
was used to make the decisions in the 
three-tier CLEO trigger system\cite{CLEOII_trigger}  
complemented by the software filter for beam-gas rejection. 
The response of the detector was modeled with
a GEANT-based\cite{GEANT} Monte Carlo (MC) simulation program.
The data and simulated samples were processed 
by the same event reconstruction program. 
Whenever possible the efficiencies were either 
calibrated or corrected for the difference 
between simulated and actual detector responses 
using direct measurements from independent data. 
Our two-photon statistics in the $\eta(1440)$ mass range 
exceeds the statistics collected by the L3 experiment 
by the factor of $\sim 5$. 
%

There are two groups of selection criteria we developed for 
this analysis. First of all we need to suppress backgrounds 
arising mainly from $e^+e^-$ annihilation to hadrons and $\tau$ pairs. 
Further, we want to select events that are reconstructed in the 
region of the detector where trigger and detection efficiencies 
are well understood and associated systematic errors are under control. 
To achieve these goals we select events with four reconstructed 
charged tracks. We require at least one charged track with 
transverse momentum exceeding 250~MeV/c. This track should 
point in the barrel part of the calorimeter. 
Also, we only use events recorded with trigger configurations 
developed for events with at least two hadrons in the final state. 
The latter three requirements select events recorded 
with well-understood trigger. 
$K_s$ candidates are identified by reconstructing 
secondary vertex radially displaced by at least two 
standard deviations ($\sigma$) from the primary interaction point. 
This secondary vertex should satisfy quality criteria 
developed and calibrated using independent data samples. 
The reconstructed mass of the $K_s$ candidate should be 
within $\sim 5\sigma$'s from its nominal value. 
Signal event candidates are required to have one such $K_s$. 
The remaining two charged tracks are identified using 
$dE/dx$ and TOF information. These measurements are used to 
form normalized ({\em i.e.} per a degree of freedom) 
$\chi^2$ which is required to be within $3\sigma$. 
There is no ambiguity associated 
with this selection for $\eta(1440)$'s two-photon kinematics 
on CLEO. Total amount of energy collected in photon-like calorimeter 
clusters that do not match with the projections of charged tracks 
should be below 100~MeV. 
Such clusters could often be present in signal events because of 
split-off effects caused by nuclear interactions of 
charged hadrons with the materials of the detector. 
Finally, we require our $K_s K^\pm \pi^\mp$ candidates 
to have total transverse momentum below 100~MeV/c. The latter two selection 
criteria are powerful in suppressing $\gamma\gamma \to K^* \bar{K}^*(\pi)$ 
backgrounds that feed down to our $K_s K^\pm \pi^\mp$ signal 
in events where some hadrons miss to be detected. 
The presence of these backgrounds in our data was proven 
by an independent analysis where we did full reconstruction of 
$K^*\bar{K}^*(\pi)$ final states arising from two-photon fusion. 

The particular criteria values used in our event selection were optimized 
to provide the best discriminating power for the signal and backgrounds 
and/or to reduce systematic uncertainty in the final result. 
The optimizations have been carried using calibration data sample 
briefly discussed below and extensive MC samples generated for this analysis. 
To measure the efficiencies of our selection criteria 
and to evaluate systematic errors we used data events where 
we identified two high-quality $K_s$ candidates. 
We used the distribution of these events' transverse momentum 
to establish their consistency 
with two-photon production mechanism. 
We measured the efficiencies of various selection 
criteria by imposing them on $K_s K_s$ data and MC samples. 
Systematic errors were estimated by comparing the efficiencies 
in data and MC simulation.  
Overall detection efficiency of our selection is 0.84\% for a hypothetical 
resonance with the mass $M=1475 {\rm ~MeV/c^2}$ and full width $W = 50 {\rm ~MeV}$. 
We chose to present the efficiency measured for these values of the parameters 
because these are the central values reported by the L3 experiment\cite{L3}. 
Relative systematic error on the efficiency is 30\%. The dominant sources 
of this error are the uncertainties in the trigger efficiency (14\%), 
four-track event selection (21\%), tight unmatched calorimeter energy (10\%) 
and event transverse momentum (10\%) requirements. 

We tested our analysis technique by estimating the two-photon partial 
width of the $\eta_c$ that we reported recently from an independent 
analysis of the same data sample\cite{etac}. We previously measured 
$\Gamma_{\gamma\gamma}(\eta_c) = (7.6 \pm 0.8 {\rm ~(stat.)} \pm 0.4 {\rm ~(sys.)}) 
{\rm ~keV}$, while our current estimate yields 
$\Gamma_{\gamma\gamma}(\eta_c) = (6.7 \pm 0.9 {\rm ~(stat.)}) {\rm ~keV}$. 
This proves that our selection criteria are unbiased. Our large systematic 
error only applies to low-mass $K_s K^\pm \pi^\mp$ final states. 

\section{Preliminary Results}
\label{results}

\begin{figure}[hbt]
\parbox{0.22\textwidth}{\epsfig{file=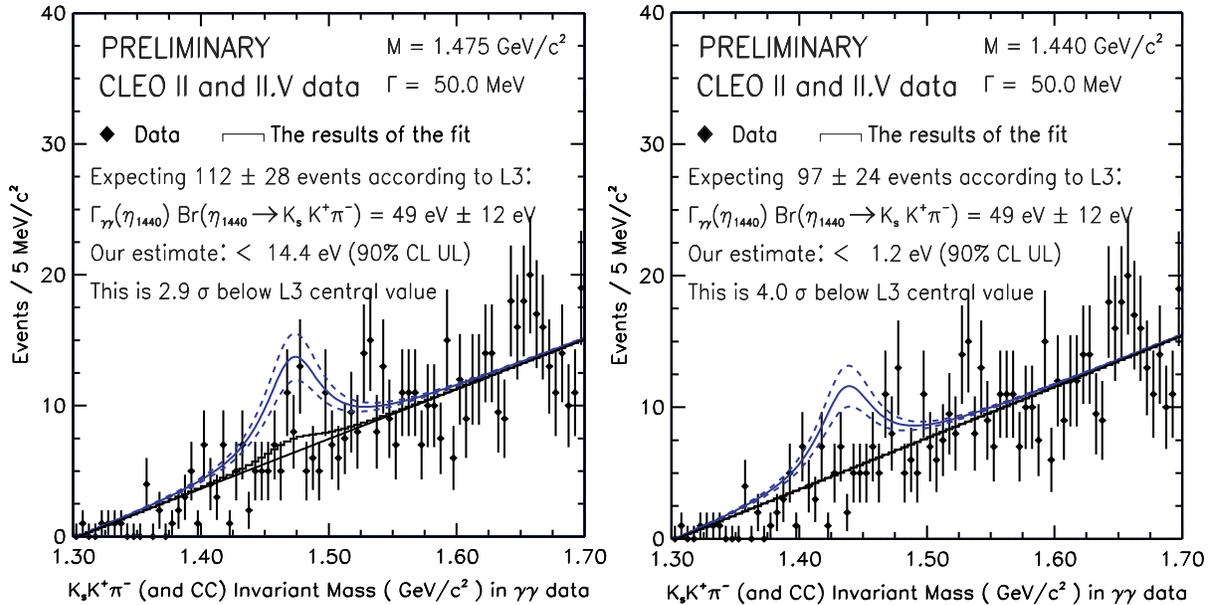,height=0.50\textwidth}} 
\caption{\label{fig_1} $K_s K^\pm \pi^\mp$ invariant mass in data, ML fit results and comparison with L3.}
\end{figure}

We show the invariant mass of $K_s K^\pm \pi^\mp$ candidates in 
data in Fig.~\ref{fig_1}. The points with the error bars show our data, 
the lines and a curve show the results of the binned maximum likelihood (ML) fit 
described below. Two dashed curves enclosing the solid curve show 
the signal expectation ($\pm 1 \sigma$ (stat.)) according to the results of 
the L3 experiment superimposed on top of background obtained from our fit. 
We observe no indication of $\eta(1440)$ or any other resonance in shown 
mass region. We conclude that our sensitivity is not sufficient to detect 
two-photon production of $\eta(1440)$ followed by its decay into  
$K_s K^\pm \pi^\mp$. To estimate the upper limit on the 
number of signal event candidates we assume that there is only 
one resonance potentially decaying into $K_s K^\pm \pi^\mp$ in the mass region 
between 1.3~${\rm GeV/c^2}$ and 1.7~${\rm GeV/c^2}$ and 
fit the distribution shown in Fig.~\ref{fig_1} (on the left) 
with signal line shape for the $\eta(1440)$ and a straight line approximating 
background contribution. There are several steps involved in 
estimating signal line shape from MC: 
first we convolute simple relativistic Breit-Wigner for a pseudoscalar 
with two-photon luminosity function. Then we convolute the resulting function 
with the detector resolution and efficiency functions. 
All these functions depend on the $K_s K^\pm \pi^\mp$ invariant mass 
and, as the result the original Breit-Wigner shape is distorted 
making the higher-mass tail of signal line shape 
to be more significant than the lower-mass tail for the results of the fit. 
The number of signal events we expect to measure assuming the values of $M$ and $W$ 
reported by L3 is $112 \pm 28$. 
However, from the results of our ML fit we estimate 90\%~CL 
upper limit on the number of signal $\eta(1440)$ events to be less than 33.  
To estimate the upper limit on the two-photon partial width of the $\eta(1440)$ 
we divide this number by the overall detection efficiency reduced by 30\% of itself, 
by the integrated $e^+e^-$ luminosity and by numerical prediction for 
pseudoscalar two-photon cross section evaluated\cite{ggmc} 
assuming $\Gamma_{\gamma\gamma}(\eta(1440))=1 {\rm ~keV}$ 
This procedure gives us 90\%~CL upper limit on the product 
$\Gamma_{\gamma\gamma}(\eta(1440)) {\cal B}(\eta(1440) \to K_s (\pi^+\pi^-) K^\pm \pi^\mp )$ 
in units of keV. First we perform this estimate for the mass and total width reported by L3, then 
we also make estimates for other values of $M$ and $W$ for a hypothetical hadron ${\cal R}$ in question. 
Some of our preliminary results for the upper limits on the value of 
$\Gamma_{\gamma\gamma}({\cal R}) {\cal B}({\cal R} \to K_s (\pi^+\pi^-) K^\pm \pi^\mp )$ 
are 14.4 eV (1475, 50), 1.2 eV (1440, 50) and 1.3 eV (1420, 20), where 
the numbers in parentheses are the values of $M$ (in ${\rm MeV/c^2}$) and $W$ (in MeV) 
used for a particular estimate. These numbers are 2.9, 4.0 and 4.0 $\sigma$'s 
below L3 measurement 
$\Gamma_{\gamma\gamma}(\eta(1440)){\cal B}(\eta(1440) \to K_s (\pi^+\pi^-) K^\pm \pi^\mp )= (49 \pm 12 {\rm ~(stat.)}) {\rm ~eV}$. 

With the data sample that exceeds the L3 statistics by the factor of $\sim 5$ 
we do not confirm their first observation of the $\eta(1440)$ 
in two-photon collisions. Our upper limit on the two-photon partial width of 
this resonance is consistent with the glueball and 
the radial excitation hypotheses that we can not rule 
out with the sensitivity of our experiment. 
We expect to receive definitive answers to the $\eta(1440)$ questions  
with our future CLEO-c experiment\cite{cleoc} 
where we plan to analyze radiative hadronic decays 
of $\sim 1$ billion $J/\psi$'s. 
We gratefully acknowledge the effort of the CESR staff 
in providing us with excellent luminosity and running conditions.


\begin{thebibliography}{9}

\bibitem{PDG} 
K.~Hagiwara {\it et al.}, Review of Particle Properties, 
{\it Phys. Rev.} {\bf D66}, 010001-493 (2002).

\bibitem{page} 
T.~Barnes, N.~Black and P.R.~Page, 
{\it arXiv} http://arXiv.org/abs/nucl-th/0208072.

\bibitem{L3} 
L3 Collab., M.~Acciarri {\it et al.}, 
{\it Phys. Lett.} {\bf B501}, 1 (2001).

\bibitem{CLEOII_description} 
CLEO Collab., Y.~Kubota {\it et al.}, 
{\it Nucl. Instrum. Methods} {\bf A320}, 66 (1992).

\bibitem{CLEOII.V_description} 
T.~Hill, 
{\it Nucl. Instrum. Methods} {\bf A418}, 32 (1998).

\bibitem{CLEOII_trigger} 
CLEO Collab., C.~Bebek {\it et al.}, 
{\it Nucl. Instrum. Methods} {\bf A302}, 261 (1992).

\bibitem{GEANT} 
R.~Brun {\it et al.}, GEANT 3.15, 
CERN Report No. DD/EE/84-1 (1987). 

\bibitem{etac} 
CLEO Collab., G.~Brandenburg {\it et al.}, 
{\it Phys. Rev. Lett.} {\bf 85}, 3095 (2000).

\bibitem{ggmc} 
M.~Budnev {\em  et  al.}, 
{\it Phys. Rep.} {\bf C15} (1975), 181.

\bibitem{cleoc} 
CLEO-c and CESR-s Working Group, 
CLNS 01/1742, \\
http://www.lepp.cornell.edu/public/CLEO/spoke/CLEOc/.

\end{thebibliography}
\end{document}